# Make search become the internal function of Internet

Wang Liang[1], Guo Yi-Ping[2], Fang Ming[3]

[1, 3] (Department of Control Science and Control Engineer, Huazhong University of Science and Technology, WuHan, 430074 P.R.China)

[2](Library of Huazhong University of Science and Technology, WuHan 430074 P.R.China)

E-mail: wangliang_f@yahoo.com

Phone: +86-27-87553494

Abstract: Domain Resource Integrated System (DRIS) is introduced in this paper. DRIS is a distributed information retrieval system, which will solve problems like poor coverage, long update interval in current web search system. The most distinct character of DRIS is that it's a public opening system, and acts as an internal component of Internet, but not the production of a company. The implementation of DRIS is also represented.

Keywords: search engine, domain, information retrieval, distributed system, Web-based service, information network

**1 Introduction**

As a customer of Internet, we usually use search engine to find information, but there are several obvious inconveniences we always encounter [1].

- There are two many unrelated records in results. We may have to turn several pages to find the real record.
- Some results are too old. The update interval of pages database is almost one month, so you need not believe the lottery number the search engine tell you.
- Every user obtains the same results for a query words from the search engine, which can't express our special interests.
- Missing much useful information. There is no a search engine which can cover all the pages on the Internet, needless to say the information in BBS, FTP or IEEE's databases.

All in all, we can't always obtain the precise and comprehensive results from current search engine.

Seeing from the whole Internet, current search engine is playing a dangerous role.

- It is ridiculous that hundreds of search engine access each site. These spiders will extract all of the pages from the web sites over and over again, which great increase the traffic load of Internet.
- Since Internet is the computer of next generation, search engine can be treated as its operation system. But when a search engine become the only dominating searches system, we may have to bear the bothersome advertisements which are added to the results that have already been not so satisfactory

We can see the origin of problem from the development of Internet. DNS-Yahoo-Google, It's the explosion of Internet that promotes the improvement of "Web information management system ". Now the Internet is still expanding. The limitations of current search engine have already become very obvious.

The other reason that results in the difficulty in information retrieval is the freedom principle of the Internet. You can release any kinds of information and not comply with any rule. But the web spider may never find such web pages if you don't register it in DNS or search engine. There are many other kinds of information on the Internet, but till now management only exists in "physical Internet ", but not in "information Internet ".

Obviously, any current search engines belong to a company can't build a database that can completely mirror the exponentially increasing pages on Internet. And they may also encounter many bottleneck problems when the size of its database reaches some critical values. So improving some internal content of Internet to meet the new demand of its customer may be an alternative method to solve current problems in information retrieval. This just is our basic idea.

As mentioned above, it's the tremendous size of Internet and the freedom principle of web information's distribution that bring the most problems in information retrieval, so the solution is evident. First, we should convert the Internet into a "smaller" one. Second, a management system should be built to administrate the distribution of information. This management system had better not give any additional restriction to the web composer.

We just propose such an information management system, DRIS (Domain resources integrated system), which truly make information retrieval become an internal function of Internet and can act as the information management system of Internet. A testbed of DIRS is also built in CERNET (China education and research network). CERNET is composed of all the networks of Chinese universities, which corresponds to the domain of "edu.cn".

**2 DRIS**

The main idea of DRIS is very simple. In DRIS, the domain (the definition in DNS) is defined as the basic cell of "information Internet" but not web server. All the web pages in one domain will be indexed and rearranged in a central server. Then all the search applications are created on these servers. The theory of DRIS will be detailed as follows.

2.1 Why domain?

The first question we should answer is that why to use the domain as basic component for information retrieval. The reasons are detailed as follows.

- On the Interne all the Web servers have already been strictly arranged by domain. So using the DNS, we can easily know how many Webs servers are in one domain, and indexing these Web pages will be more easily.
- Normally, each domain corresponds to an organization. For instance, "hust.edu.cn " is our university's domain name. All the servers under this domain and a DNS server are managed by us. Since the DRIS server will extract and index all the pages in this domain, the organization that corresponds to this domain should have authority on own their DRIS and can efficiently control the distribution of information. So we answer a crucial question, that is who is in charge of the DRIS server.
- If the DRIS is brought to realization, DNS can be easily updated to DRIS.

2.2 The architecture of DRIS

Since the definition of domain is used, the architecture of the DRIS is as same as that of DNS. It's shown in figure 1.

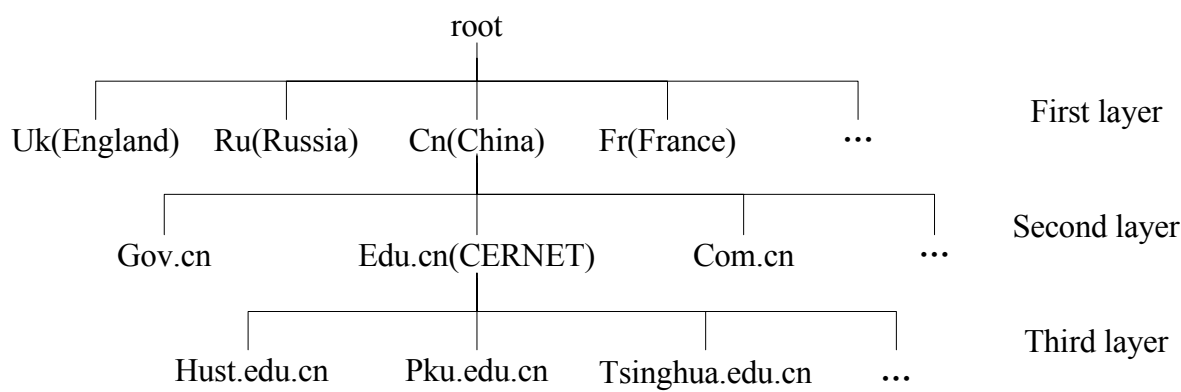

Figure 1  The architecture of the DRIS

The node in the third layer corresponds to a third level domain like a university, and the second layer may be the sub network of one country like domain of "edu.cn". The node in top layer normally is a country except USA. DRIS is a hierarchical distributed information retrieval system. Every node of DRIS is an integrated search engine, but has different search scope.

We can always see that many sites don't comply with the discipline of DNS and use the top domain name such as "net ", "com" without adding the domain name of their country. This condition will be discussed in implementation part of DRIS.

2.3 search system in each layer

We will introduce each level's search system from third layer to the top layer.

2.3.1 Third layer: extract and Index the pages

The node in this layer just acts as a normal search engine. Only difference is that the search scope is limited to a third level domain, like a university. This search system comprise of three parts: page-fetching mechanism, indexer, and searching interface. The three parts are introduced respectively.

2.3.1.1 Extract the pages

Page-fetching mechanism will download all the pages in a three level domain. For example, "[www.hust.edu.cn](www.hust.edu.cn) " is the domain name of our university, so the server under the domain of "hust.edu.cn" like the department of CS's server "cs.hust.edu.cn " can be easily found by referring to the DNS server. Then the spiders can download all sites included in this domain.

The work of spider is organized by single web site. When a spider visits a web server, it will download all the content of this site and stop working when it encounters a URL that links to other site. The content in this kind of URL in other site is also treated as valuable matter of this site and be download either. This is because such links also express the interest of this site, and it's important when we rank the pages by using the Hyperlink Analysis technology [2] in the following parts.

2.3.1.2 Index the pages

Now there is still no a "standard index technology" for the search engines. Normally, the key issue in indexing is the appreciate selection of metadata. In the common sense, a web page is represented by the keywords with its ranking score. We also use this method to index the web pages. By this means, the rank score of keyword will be determined by the position and frequency that the word appears in the whole text. The other fields such as title, encoding and abstract will also be used to describe a web page.

2.3.1.3 Search interface

Providing user interface and processing the search results is main function of search interface. Ranking method is the key issue to search interface. Now there are two methods to rank the pages. First, the score of keyword. In this method, the score of keywords [3] is calculated by the frequency and the place of this word appearing in the page. When a query word is submitted, all the pages that contain that word will be retrieved and then ranked by the score of corresponding keywords. The other method, Hyperlink Analysis, is applied by some current search engines, like Google [4]. By this means, the rank score of a page is determined by the number of link that point to this page.

In this layer, we use the fist method to rank the results. This is because in this condition, the pages are limited to a small area like a school, but the Hyperlink Analysis is more useful in a large scale of Internet. So score of keywords is enough.

2.3.2 Second layer: Harvest the metadata

This layer will provide the search in all the servers under a second levels domain. The search engines of this layer have only two parts, web page database and search interface but no its own spider. The data of search engine is downloaded directly from the database of the search engine in the corresponding third layer. For example, the search engine corresponding to the domain "edn.cn" will obtain its data from the page databases in different schools in China.

The search engine in this layer only harvests the metadata from the pages databases in the third layer, but not all the pages itself. In this method, the update interval can be shorter than directly extracting web pages from thousands of web servers.

A notable problem that should be concerned is the overlap of web pages. In the third layer, when the spider extracts pages from a site, it also gets some pages which don't belong to this site. So when harvesting the metadata, some pages may appear many times. As the download method in the third layer, this number is the number that the other sites direct to this page, except the page itself. According to the theory of Hyperlink Analysis, this number will be treated as the ranking score of this page.

As mentioned above, Hyperlink Analysis will be applied to rank pages in this layer, which is different from the ranking method in the third layer. This is because the scope in this level's domain, like all the universities in China, is large enough and the Hyperlink Analysis becomes efficient in this condition.

2.3.3 Top layer: a distributed search system

The two former layer's search engines are all centralized structure, but the search engine in this layer will be distributed information retrieval system. It has merely one part, search interface, no spider, and no page databases.

There are three issues when designing a distributed search system [5].1 the underlying transport protocol for sending the message (e.g., TCP/IP). 2 a search protocol that specifies the syntax and semantics of message transmitted between client and servers.3 a mechanism that can combine the results from different servers. These problems are detailed as follows.

1 Communication protocol. In DRIS, SOAP is applied as the basic protocol for communication, because the SOAP has many advantages than HTTP in security, opening, etc.

2 search protocol. The search protocol of DRIS is based on Webservice. Webservice uses the SOAP as the fundamental protocol and provides an efficient distributed structure. We refer to the SDLIP [6] and Google's search Webservice [7] to define the format query and results. By this means, all the search engines in the second layer should provide the standard search Webservice according to this protocol. And Search engine in this layer just needs to index all the Webservice in lower layer.

3 combining the results. The key problem for the combination of results is the rank of pages from different servers. In the second layer, the ranking score of one page is represented by its overlap number in databases. In this layer, we also use this theory to calculate the rank of document. The score of the same page from different database will be simply added up and a finial ranking list of documents will be produced. Some other factor like the speed of different service will be also considered in ranking the results in the further work.

The work theory of this search engine is just like that of Meta search engine, which has no its own pages database, but an index of the search interface of other engines. But the sub search engine of DRIS is arranged strictly, complied with the same protocol, so performance of this Meta search system is much better than any current Meta search engine. The search engine in this layer will provide the search in a country, which will cover the most of search request on Internet. If you want to search parallel in several countries, you can use the search Webservice in top layer to build a special search engine. The problem of multiple languages processing should be considered in this application.

2.3.4 The OO (Object Oriented) module of DRIS

All the nodes in DRIS will provide two kinds of search interface, user interface based on web and application program interface (API) based on Webservice. The web interface is just as same as normal search engine. But there are some special rules in search Webservice of DRIS for the advantage of API user. The search function and its interface are all the same in different layers' search Webservice. And all the Webservice in DRIS will be arranged by the Object Oriented theory.

As an Object Oriented software system, we use the class tree to describe the whole system. All the nodes in the DRIS are arranged under the namespace "DRIS" and be treated as the child class of DRIS. All search Webservice should provide the service through URL of "DRIS. Domain name" and the main class name of this Webservice will be "DRIS. Reversed domain name". For example, the domain name of our university is "hust.edu.cn". Then our DRIS server will provide the search service through link of "DRIS.hust.edu.cn" and its main class name is "DRIS.cn.edu.hust ". The class tree of DRIS is shown in fig2.

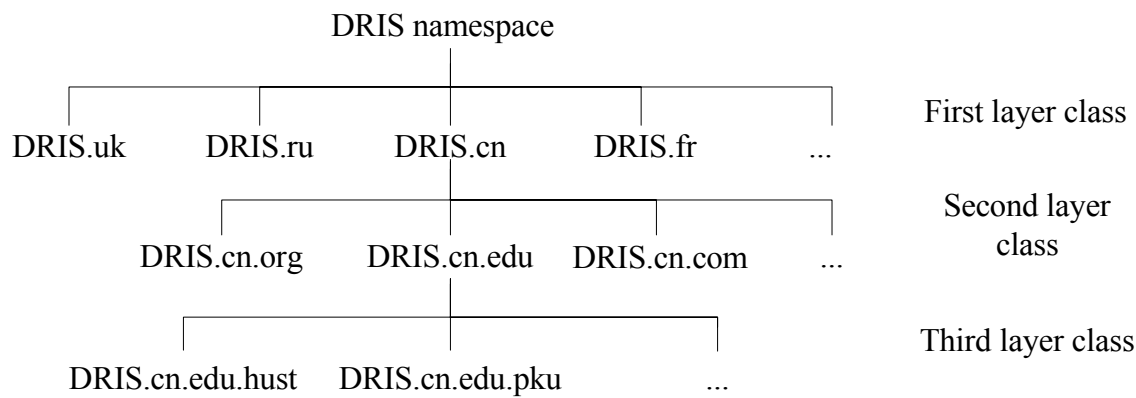

Figure 2 the class tree of DRIS

This naming system gives many conveniences when applying search service in DRIS. First, if you want to use the search service of a node, just knowing its domain name is enough. Second, a higher node will index all search Webservice under its domain to maintain the inherit relation between layers. So you can only cite a Webservice of a node and use all the search service under its domain. For example, you can cite the service from "DRIS.edu.cn" in your program, and when you want to visit the search service the in our university, "hust ", you can directly use the function in child class "DRIS.cn.edu.hust" and need not cite it.

## 3 The advantage of DRIS

Normally there are three principles to judge a search engine.

1. The coverage of the search engine. The more pages are indexed, the better a search engine is. In fact, the DRIS can embody almost all the pages on the Internet.
2. The update speed of search engine. All the work of downloading the pages is carried out by many distributed spiders in the third layer of DRIS, so the content of search engine's databases can be updated everyday.
3. Quality of search service. Intelligent personal search service will be the general trend to provide a better search service. Now the data source of current intelligent search system comes from two places, pages database built by them or using the search service of current search engine, like Google's API. These two methods both have obvious disadvantages. On the contrary, DRIS provide standard search service in three levels' scope of Internet and is free. From our point of view, personal intelligent search engine will be the most promising service based on DRIS.

All in all, DRIS first makes the size of Internet "smaller ". The DRIS servers will be the basic component for information retrieval. And all the pages on Internet are rearranged in a uniform format by DRIS servers. So as the designer of search engine, his spider need not visit billions of diversiform web pages, but thousands of normative pages databases.; as the web composers, they can still use HTML or other format to publish their information as their wish; as the customer of Internet, you can truly get the results you really want, which is more comprehensive and precise than before.

## 4 the implementation of DRIS

Although the theory of DRIS can make the information retrieval more efficiently, but when taken into realization, the most important thing that we should consider is why the organizations willing to build the DRIS server, and what they can benefit from this system. Just indexing their resources for the advantage of others may not be a sufficient reason to build such a system. DRIS should first solve some urgent problems of their own and then benefit the others. We fully consider this condition when building testbed of DRIS in CERNET. The implementation procedure of DRIS is introduced as follows.

4.1 union search systems in single university

First we build a third layer DRIS system in the university, which is the underlying structure of DRIS. And fortunately, we have found some urgent need to build such a system. The reason to build a sub DRIS is introduced as follows.

- A university needs a Web search engine that indexes all the pages in school. Now many universities haven't built their own search engine, though there may be hundreds of web servers in the school. If you want to find some notes, you may have to remember many URL of sites. Now many sites use the Google to execute the search in site, but we always need to search all the sites in school .If such a standard excellent and free search engine is proposed, I think many universities will adopt it.
- A university needs Union search system for all resources. In fact, there are many other resources in school besides web pages, such as FTP, many digital resources in library and each department, but there no a mature system to manage all these information. Even just in the library, hundreds of databases have been introduced these years, and we always have to search in different search interfaces one by one to get a comprehensive result. Everyone hope to "search in one page, and then search all the school ".

Firstly, the web search engine for a university is introduced. The standard DRIS only indexes the servers under the corresponding domain, but in fact many servers are not strictly named by the university's domain and some sites even haven't a domain name. So we should first determine how many servers in the school. Most of servers have been registered in the network center and we just need to get a list from there, but we still scan all the IP of our school to ensure all the servers are indexed. So in practice, the "domain" in DRIS is always a sub network of Internet, not strictly corresponding to the "domain" in DNS, but the naming method of Webservice has no change.

Then a union search engine that can integrate many kinds of resources is proposed. Information retrieval for heterogeneous resources is still an unsolved problem. Some digital library projects like InfoBus just deal with this issue. The key to the search in the heterogeneous resources is an efficient middleware, the SDLIP and Z39.50 is just such protocol to build the middleware. After studying these protocols, we propose our experimental protocol [8] based on SOAP/Webservice/UDDI. This protocol can be applied by web search engine, professional databases, FTP, etc. Based on this protocol, we have built a union search system that can search heterogeneous resources in our school. In this system, all data source should provide the uniform search Webservice. And if one system can't provide such an interface, our systems will automatically transform its search interface into standard one. The user can search many kinds of information like web pages, file in FTP, and digital book in one search interface.

All the search services in our university are also arranged by Object Oriented module. The class tree of our university's node in DRIS is shown in fig3.

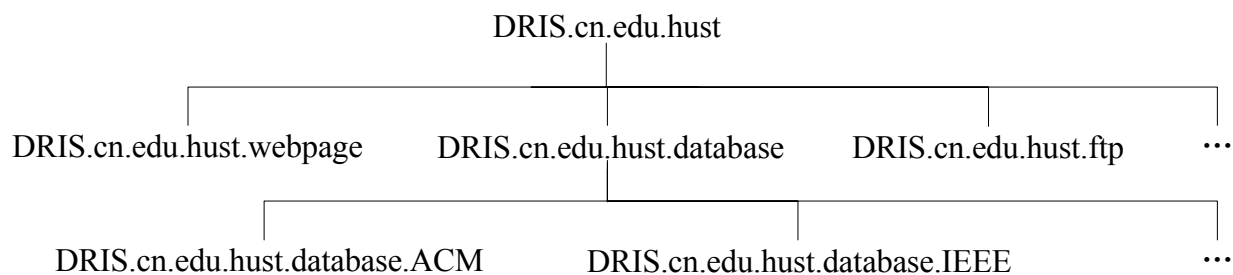

Figure 3 the class tree of DRIS.cn.edu.hust

All the nodes will provide standard search Webservice. Top class "DRIS.cn.edu.hust" will provide a union search

Webservice of all the resource in our school. The function in the class of "DRIS.cn.edu.hust.webpage" will provide the web pages search service. "DRIS.cn.edu.hust.ftp" and other class will also provide corresponding search service. Child class of "DRIS.cn.edu.hust.database" will provide the search of many professional databases of our university, such as the digital resource purchased by library.

4.2 search system in CERNET

Merely as a Web search engine, our search engine can distribute freely to other universities. Then the web search system in second layer of DRIS can be easily built. Search system in CERNET will integrate all the web pages in Chinese universities.

But what most attractive to build second layer of DRIS is not the web page search engine, but the characteristic resources in different schools. Till now, there are no a mature interoperation protocol for the exchange of information in different organization. The Protection of copyright and security consideration also bring some troubles in exchange. A protocol that can deal with these problems should be proposed in the future. The data fusing and resources selection should be especially considered in this protocol.

4.3 search systems in China

If all search systems in the second layer are finished, just as in CERNET. The search engine of a country could be easily created. As the final version, DRIS will integrate not only web pages, but also the other kinds of resources. More issues in this layer will be considered in the further experiment.

**5 further work**

Main work in the future will be concerned with the standard protocol of DRIS, and more experiment is needed to improve such a protocols. Some hardware resources can also be integrated by using the basic idea of DRIS. Now some Grid Computing technology like Globus is just applied in a small area, a large-scale application in the whole Internet may be more useful. DRIS will integrate all the resource on internet and turn Internet into a supercomputer. If these assumptions come into realization, you can obtain all the information of this world and powerful computing ability from the Internet from your PDA.